\documentclass[a4paper,11pt]{article}
\usepackage{jinstpub} 
\usepackage{siunitx}
\usepackage{mathtools} 
\usepackage{physics}
\usepackage{lineno}

\title{\boldmath Fiber-coupled Digital Photo Sensors for Large Time Projection Chambers}

\author[a]{P. A. Breur}
\author[a]{, X. Defay}
\author[c]{, D. Jackson}
\author[a,1]{, J. R. Sinclair\note{Corresponding author.}}
\author[b]{, E. Triller}
\author[a]{, and A. Young}

\affiliation[a]{SLAC National Accelerator Laboratory,\\ Menlo Park, CA, USA}
\affiliation[b]{Michigan State University,\\ East Lansing, MI, USA}
\affiliation[c]{College of the Desert,\\Palm Desert, CA, USA}

\emailAdd{jimsin@slac.stanford.edu}

\abstract{This paper presents a novel approach to addressing challenges in neutrino event reconstruction within large Time Projection Chambers (TPCs). 
By integrating fiber-coupled digital silicon photomultipliers, we propose a design that enhances light detection, improves energy resolution, and enhance event reconstruction. 
Advancements in power and signal over fiber technologies are leveraged to deploy digital sensors within the TPC bulk volume, enabling precise timing and robust particle identification.}


\begin{document}
\maketitle
\flushbottom

\section{Introduction}
\label{sec:intro}
This work describes a detector concept based on recent advancements in photonics and digital Silicon PhotoMultipliers (dSiPMs). 
Large Liquid Argon Time Projection Chambers (LArTPCs) are finding application in neutrino oscillation experiments, such as the Deep Underground Neutrino Experiment (DUNE). 
GeV-scale neutrino reconstruction is challenging due to Deep Inelastic Scattering (DIS) and multiple hadron interactions.
Scaling TPCs increases optical paths beyond the Rayleigh scattering length, limiting performance. 
dSiPMs in combination with photonics have the potential to address these drawbacks.
This concept focuses on measuring prompt scintillation light to address these challenges.

\section{Physics Motivation for utilizing prompt scintillation}
Fast timing in TPCs is crucial for multiple physics applications.
The multiple-GeV energies required for neutrino oscillation measurements will cause DIS~\cite{alex}.
DIS produces multiple hadrons and re-interactions in the final state. 
This complicates reconstruction as scattering hadrons make the primary vertex non-obvious. 
Energy reconstruction is also biased by neutron production as neutrons can carry away a significant amount of energy from the vertex or if a shower produces a neutron. 
Fast neutrons typically cause a proton or nuclear recoils within a few nanoseconds. 
Neutron energy reconstruction through time of flight may be possible with fast timing information; a \SI{100}{\mega\electronvolt} neutron travels \SI{12}{\centi\meter\per\nano\second}.

Another important application of fast timing is Particle IDentification (PID) through timing. 
Consider, for example, searches for the proton decay channel $p\rightarrow K^++\overline{\nu}$. 
The $K^+$ decays with a lifetime of \SI{12.4}{\nano\second} to $\mu^+ + \nu_\mu$ with a 64\% branching ratio~\cite{pdg}. 
The charge signature of proton decay is complicated by the various $K^+$ decay modes and strong interactions with residual nucleus~\cite{ChrisKaons}, making the charge signature difficult to distinguish from that of atmospheric neutrinos. 
Scintillation light is produced by the initial signal, and then by the $K^+$ decay products. 
The $K^+$ lifetime can be used for PID by reconstructing the optical signal with $\order{\SI{1}{\nano\second}}$ timing resolution, as in Super-K~\cite{SuperKKaons}. 
Similarly, PID with timing would also help with $\pi\rightarrow\mu$, $K^0_L\rightarrow X$, etc. as long as the lifetime is over a few nanoseconds.

To observe these lifetimes, measure fast neutron energies, or reliably identify the neutrino vertex in a DIS interaction we must have access to prompt scintillation light.  

\section{Scintillation light in Argon}
Argon scintillation emits photons from singlet and triplet excited states with lifetimes of $\order{\SI{1}{\nano\second}}$ and $\order{\SI{1}{\micro\second}}$, respectively~\cite{scintillation}. 
Slow scintillation is unsuitable for nanosecond-scale event reconstruction. 
Detection of prompt light is hindered by Rayleigh scattering, which smears out photon arrival times by $\order{\SI{10}{\nano\second}}$ for propagation distances of $\order{\SI{1}{\meter}}$~\cite{Rayleigh}. 
To have access to prompt light, optical paths must be less than the Rayleigh scattering length, and the sensor response must be $\order{\SI{1}{\nano\second}}$ or better.
By instrumenting the TPC volume with IceCube-esque light readout~\cite{IceCube} and charge readout, we aim to utilize prompt light for enhanced event timing and PID.

\section{Proposed Detector Design}

\begin{figure}
    \centering
    \includegraphics[width=0.7\linewidth]{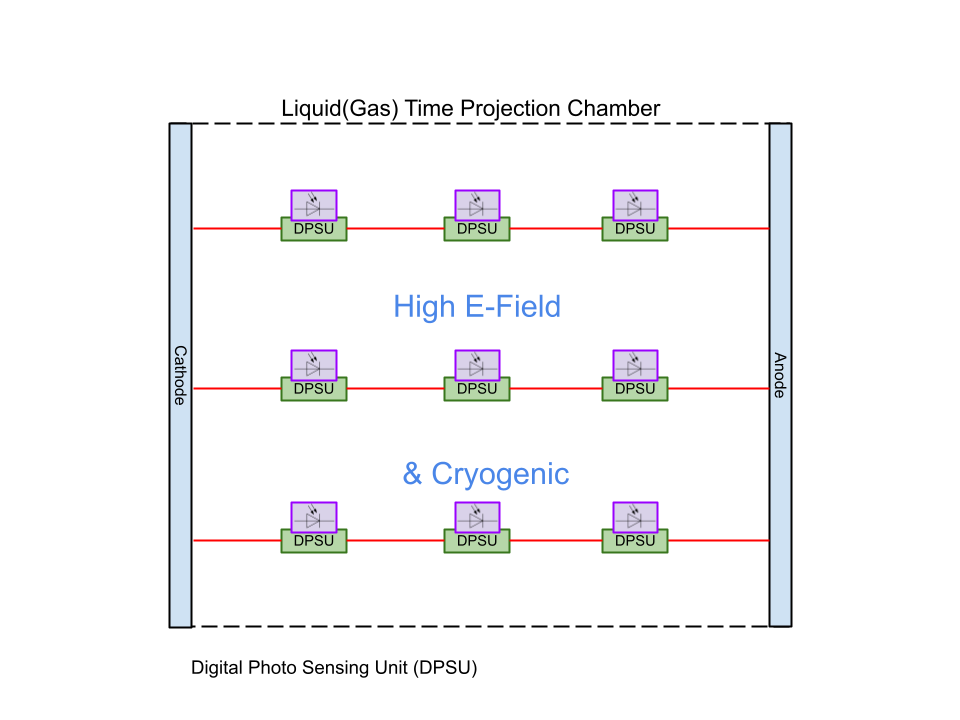}
    \caption{Arrays of Digital Photo Sensor Units (DPSU) inside a Time Projection Chamber. Each DPSU consists of photo sensors and associated readout electronics with an optical power converter and silicon photonics. DPSUs are dielectrically coupled with optical fibers to enable deployment inside electric fields.}
    \label{fig:DPSU}
\end{figure}

The detector concept is based on dSiPMs in conjunction with Power over Fiber (PoF) and Signal over Fiber (SoF).  
By dielectrically coupling the dSiPMS with optical fibers they can be deployed in arrays of Digital Photo Sensor Units (DPSUs) inside the TPC between the cathode and the anode, as illustrated in Figure~\ref{fig:DPSU}.

Each DPSU integrates dSiPMs, readout electronics, and optical power converters and silicon photonics for signal transmission.    
dSiPMs are ideal for this, as precise threshold-crossing times obviates the need for waveform analysis, and reduced data acquisition complexity. Their digital output also enables higher channel density and simplified data transmission through digital multiplexing. 

\begin{figure}
    \centering
    \includegraphics[width=0.5\linewidth]{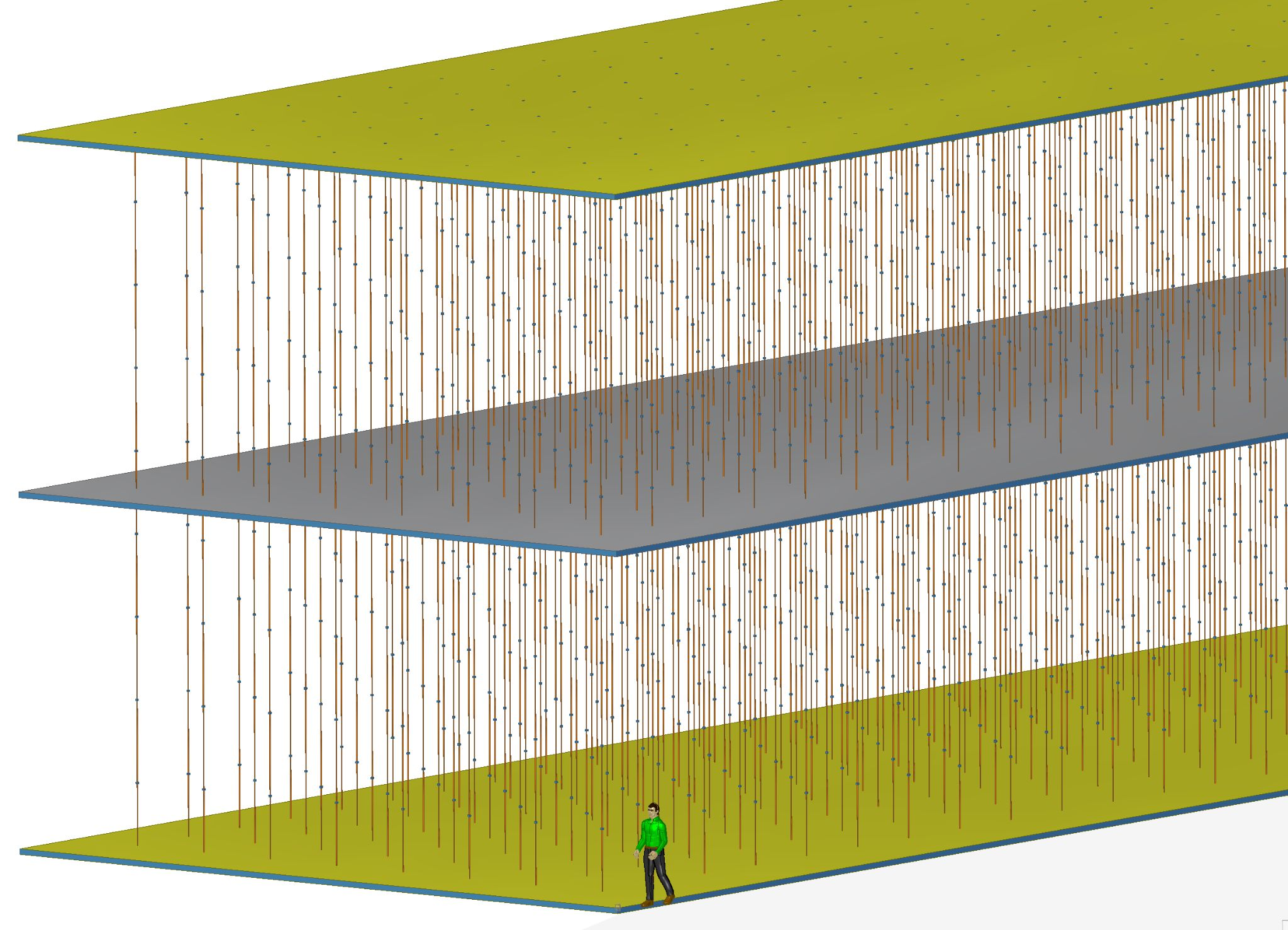}
    \caption{Arrays of Digital Photo Sensor Units (DPSUs) inside the DUNE vertical-drift geometry \numproduct{60x13x13.5}~\unit{\metre\cubed}~\cite{VD}. 
    39 strings of DPSUs are separated at 1.5 times the Rayleigh scattering length, with 8 DPSUs along each string.}
    \label{fig:VD}
\end{figure}

Consider application to a detector of the same dimensions as DUNE's vertical-drift geometry: \SI{60}{\meter} long, \SI{13}{\meter} wide, and \SI{13.5}{\meter} tall~\cite{VD}. 
To ensure the optical path is less than a scattering length, strings of sensors are displaced at 1.5 times the Rayleigh scattering length, \SI{1.5}{\meter}, with the same separation is applied along each string. 
The separation of and coverage at each DPSU will be further optimized through a more detailed study; it will strongly depend on the achievable photon detection efficiency. 
Yielding an $8\times39$ grid of strings with 8 sensors on each string, totaling only 2500 DPSUs for the entire volume, as shown in Figure~\ref{fig:VD}.
For comparison, the Sudbury Neutrino Observatory~\cite{sno} is a \SI{12}{\meter} diameter sphere, which used nearly 10000 photomultiplier tubes.

Unlike conventional light readout systems which occupy the periphery of the detector volume, and have access to prompt timing information from only a small volume, this approach allows access to the entire volume.
It enables localized triggers and improved energy reconstruction through efficient prompt-light detection and robust particle tagging, while providing resilience to nitrogen contamination~\cite{N2} that suppresses the slow light.

\section{Key Technological Components}

The key technologies for this detector concept are PoF, SoF, and dSiPMs.
Recent advancement for PoF in LAr have enabled light readout at the periphery of the TPC~\cite{pof}. 
However, current systems demonstrate approximately 50\% efficiency, which will require optimization to reduce heat load.
For SoF, fiber-coupled diodes have been cryogenically demonstrated, however a more energy efficient solution is to use external lasers with data transmission via modulation in silicon photonics~\cite{sof}.
There are multiple significant advancements in dSiPMs~\cite{FNALsSiPM,DESYdSiPM,3dsipm2} but typically sensitivities peak above \SI{400}{\nano\meter} without an additional wavelength shifter. 
To avoid time smearing associated with wavelength shifting, development of Vacuum UltraViolet(VUV) sensitive single photon avalanche diodes (SPADs) is needed.  

\begin{figure}
    \centering
    \includegraphics[width=0.4\linewidth]{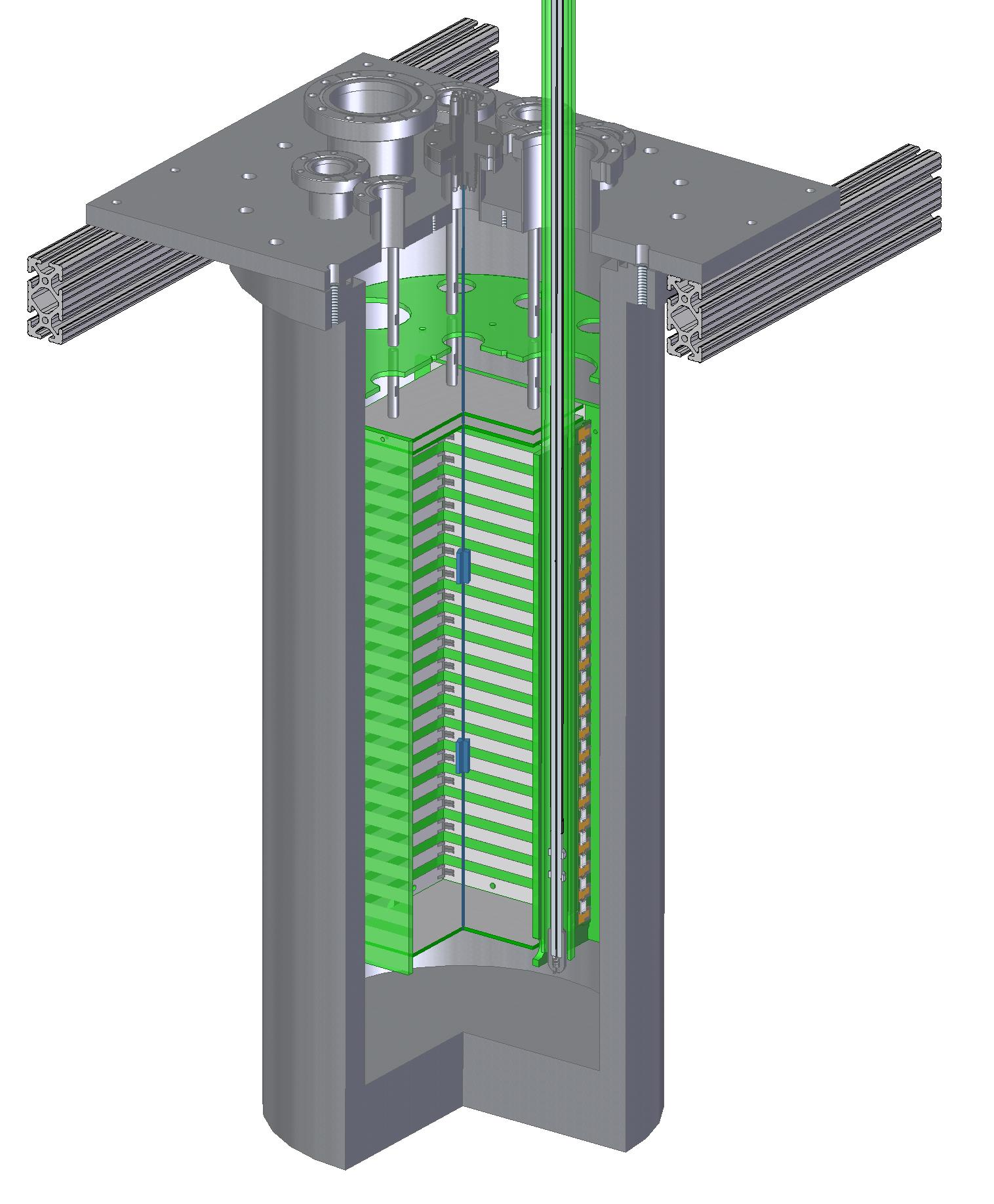}
    \caption{A \SI{50}{\centi\meter}-drift liquid argon time projection chamber currently under construction at SLAC for demonstrating fiber-coupled digital photo sensors in electric fields.}
    \label{fig:test}
\end{figure}

At SLAC, we are currently working on a developing a pulsed laser source for PoF to mitigate heat input through reduced duty cycles, in combination with storing energy in a capacitor.
As the readout is needed only when triggered, continual power is not necessary, therefore a capacitor can charged at reduced duty cycle and drawn from as required. 
A \SI{50}{\centi\meter}-drift PoF demonstration TPC is being constructed, as shown in Figure~\ref{fig:test}, using FBK’s SuperEllen~\cite{FBK} sensors as the initial digital sensor.   
In parallel, there is a research effort to enhance VUV-sensitive SPADs, leveraging SLAC’s expertise in cryogenic electronics and SPADs for light detection and ranging.

\section{Summary}
The integration of fiber-coupled dSiPMs represents a significant advancement for large LArTPCs.
This approach enables precise timing resolution, robust particle identification, and improved energy reconstruction, addressing key challenges in neutrino event detection.

By leveraging PoF and SoF technologies, this design allows for sensor deployment inside the TPC bulk volume without introducing electrical noise or compromising detector integrity. 
The proposed configuration ensures efficient light collection while minimizing scattering effects, making it a scalable and resilient solution for next-generation neutrino experiments.

Ongoing developments, such as optimizing PoF efficiency and enhancing VUV-sensitive SPADs, are critical steps toward full-scale implementation. 
We invite collaboration from researchers partners to refine these emerging technologies, explore additional applications, and advance the next generation of high-energy physics detectors.

\acknowledgments

SLAC is managed by Stanford University under DOE/SU Contract DE-AC02-76SF00515.

\bibliographystyle{JHEP}
\bibliography{biblio}

\end{document}